%% file: eprint.tex
\documentclass[12pt]{revtex4}

\newcommand\pubnumber{RESCEU-4/14,RUP-14-2}
\newcommand\pubdate{\today}

\def\Title#1{\begin{center} {\Large #1 } \end{center}}
\def\Author#1{\begin{center}{ \sc #1} \end{center}}
\def\Address#1{\begin{center}{ \it #1} \end{center}}

\newcommand\pubblock{\rightline{\begin{tabular}{l} \pubnumber\\
         \pubdate  \end{tabular}}}
\newenvironment{Abstract}{\begin{quotation}  }{\end{quotation}}
\newenvironment{Presented}{\begin{quotation} \begin{center} 
             PRESENTED AT\end{center}\bigskip 
      \begin{center}\begin{large}}{\end{large}\end{center} \end{quotation}}
\def\Acknowledgements{\bigskip  \bigskip \begin{center} \begin{large}
             \bf ACKNOWLEDGEMENTS \end{large}\end{center}}
             
             \def\dhc{\bar{\delta}_\m{hc}}
\def\dhc{\bar{\delta}_\m{hc}}
\def\Ki{K_\m{i}(r)}

\def\al{\alpha}
\def\be{\begin{equation}}
\def\ee{\end{equation}}
\def\ep{\epsilon}

\def\fr{\frac}
\def\de{\delta}
\def\ga{\gamma}
\def\de{\delta}
\usepackage{amsmath,amssymb}
\usepackage{comment}
\def\al{\alpha}
\usepackage[dvipdfmx]{graphicx}
\def\be{\begin{equation}}
\def\ee{\end{equation}}
\def\ep{\epsilon}

\def\fr{\frac}
\def\de{\delta}

\def\ga{\gamma}
\def\Ga{\Gamma}
\def\de{\delta}

\def\al{\alpha}
\def\si{\sigma}

\def\m{\mathrm}

\def\ri{r_\m{i}}

\input econfmacros.tex

\begin{document}
\begin{titlepage}
\pubblock

\vfill
\Title{Investigating formation condition of primordial black holes for 
generalized initial perturbation profiles}
\vfill

\Author{Tomohiro Nakama}
\Address{Department of Physics,
Graduate School of Science,\\ The University of Tokyo, Bunkyo-ku,
Tokyo 113-0033, Japan}

\Author{Tomohiro Harada}
\Address{Department of Physics, 
Rikkyo University, Toshima, Tokyo 175-8501, Japan}

\Author{A.~G.~Polnarev}
\Address{Astronomy Unit, School of Physics and Astronomy, \\
Queen Mary
University of London, \\ Mile End Road, London E1 4NS, United Kingdom}

\Author{Jun'ichi Yokoyama}
\Address{Research Center for the Early Universe (RESCEU),\\
Graduate School of Science, The University of Tokyo, \\ Bunkyo-ku,
Tokyo 113-0033, Japan}
\Address{Kavli Institute for the Physics and Mathematics 
of the Universe (Kavli IPMU), WPI, TODIAS,\\
The University of Tokyo, Kashiwa, Chiba 277-8568, Japan}

\vfill
\begin{Abstract}
Primordial black holes (PBHs) are an important tool in cosmology to probe 
the primordial spectrum of small-scale curvature perturbations that reenter
the cosmological horizon during radiation domination epoch. 
We numerically solve the
evolution of spherically symmetric highly perturbed configurations to clarify the
criteria of PBHs formation using a wide class of curvature profiles characterized by
five parameters. 
It is shown that formation
or non-formation of PBHs is determined essentialy by only two master parameters. 
\end{Abstract}
\vfill
\begin{Presented}
The 10th International Symposium 
on Cosmology and Particle Astrophysics (CosPA2013)\\
Honolulu, Hawai'i,  November 12--15, 2013
\end{Presented}
\vfill
\end{titlepage}
\def\thefootnote{\fnsymbol{footnote}}
\setcounter{footnote}{0}

\section{introduction}
It is well known that a region with large amplitude curvature profile
 can collapse to a primordial black hole (PBH)
 \cite{Zel'dovich-1974,Hawking:1971ei}. 
PBHs are formed soon
 after the region enters the cosmological horizon during the
 radiation-dominated epoch. 

Even if PBHs would never have been detected, existing observational constraints \cite{Carr:2009jm} 
will provide valuable information on inflationary cosmological models. 
It is important to probe 
the perturbation spectrum on significantly smaller scales as well in order to obtain
more helpful information to single out the correct 
inflation model. 

Originally the problem of PBH formation was studied analytically \cite{Carr:1974nx,Carr:1975qj}:
\be
\fr{1}{3}\lesssim \dhc,\label{classical}
\ee
where $\dhc$ is the energy density perturbation averaged over the 
overdense region evaluated at the time of horizon crossing. 
This criterion has long been used in papers on
theoretical 
prediction of PBH abundance
(but has recently been refined in \cite{Harada:2013epa}).
In this simple picture, 
the dependence on the 
profile or shape of perturbed regions has not been taken into account.

However, recent numerical analyses have shown that the condition for PBH formation does
 depend on the profile of perturbation
 \cite{Shibata:1999zs,Polnarev:2006aa} (see also \cite{1979STIN...8010983N,PhysRevD.59.124013}). 
Both \cite{Shibata:1999zs} and \cite{Polnarev:2006aa}(hereafter PM) used
two-parameter families of the initial profile and obtained two parametric conditions of PBH formation.  
It was clear from the above publications that one parametric description was not sufficient. 
However it was not clear whether the two-parametric description is good enough.
In the present paper, we extend these preceding analyses by making many more numerical
computations of PBH formation based on the initial curvature profile including many more parameters,
adopting the five-parameter family of profiles. We show that the criterion
 of PBH formation can still be expressed in terms of two
 crucial 
(master) parameters, 
even though the considered profiles belong to
the five-parametric
 family.

\if
They introduced some functions to model primordial perturbed regions 
and obtained conditions for PBH formation for those perturbations
represented by their functions. 
Though there results were new and important, there seems to be some room 
for improvements. 
Firstly, physical interpretation of their condition of PBH formation is 
not clear and therefore 
what kinds of physical mechanisms play important roles in the process of 
PBH formation can not be understood. 
Secondly, they used functions which include at most two parameters and therefore
types of initial configuration shapes investigated are limited. 
Therefore their condition for PBH formation is applicable only to 
limited types of perturbation shapes. 
In reality, various kinds of perturbations must have been produced during inflation so 
in this paper we consider a wider class of shapes of perturbations 
by intoducing a function which includes five 
parameters, which can express a far more variety of perturbation shapes, 
thereby enabling more realistic analysis 
of PBH formation condition. for this extended class of shapes, we found 
a condition for PBH formation, which is expressed by two quantities 
characterizing profiles of perturbations and whose physical
interpretation 
can also be provided. 
The condition we obtained is a lot more general and accurate than 
those obtained before. 
\fi

\section{Setting up the initial condition}
The metric used can
be written in the form used by Misner and Sharp \cite{Misner:1964je}:
\begin{equation}
ds^2=-a^2dt^2+b^2dr^2+R^2(d\theta^2+\sin^2\theta d\phi^2),\label{1}
\end{equation}
where $R$, $a$ and $b$ are functions of 
$r$ and the time coordinate $t$. We consider a perfect fluid with the energy density $\rho(r,t)$ and pressure 
$P(r,t)$ 
and a constant equation-of-state parameter $\ga$, $P(r,t)=\ga \rho(r,t)$.
We express the proper time derivative of $R$ as
\begin{equation}
U\equiv \frac{\dot{R}}{a},\label{2}
\end{equation}
with a dot denoting a derivative with respect to $t$.

We define the mass, sometimes referred to as the Misner-Sharp mass in the literature, within the shell of circumferential radius $R$ by
\be
M(r,t)=4\pi\int^{R(r,t)}_0\rho(r,t)R^2dR.\label{defofM}
\ee
We consider the evolution of a perturbed region embedded in a 
flat Friedmann-Lemaitre-Robertson-Walker (FLRW) Universe with metric
\be
ds^2=-dt^2+S^2(t)(dr^2+r^2d\theta^2+r^2\sin^2\theta d\phi),
\ee
which is a particular case of (\ref{1}). The scale factor in this background evolves as
\begin{equation}
S(t)=\left(\fr{t}{t_\mathrm{i}}\right)^{\alpha} ,\quad\alpha\equiv\frac{2}{3(1+\gamma)},
\label{S0anddefofalpha}
\end{equation}
where $t_\mathrm{i}$ is some reference time.

The background Hubble parameter is
\begin{equation}
H_0(t)=\fr{\dot{R_0}}{a_0R_0}=\frac{\dot{S}}{S}=\frac{\alpha}{t}.
\end{equation}
The energy density perturbation is defined as
\be
\de(t,r)\equiv \fr{\rho(t,r)-\rho_0(t)}{\rho_0(t)}.\label{defofdelta}
\ee
The curvature profile $K(t,r)$ is defined by rewriting $b$ as
\be
b(t,r)=\frac{R'(t,r)}{\sqrt{1-K(t,r)r^2}}\label{16}.
\ee
This quantity $K(t,r)$ vanishes outside the perturbed region so that the solution asymptotically approaches the background FLRW
 solution at spatial infinity.

We denote the comoving radius of a perturbed region by $r_\mathrm{i}$, the 
precise definition of which will be given later (see eq. (\ref{defri})), 
 and define a dimensionless parameter
$\ep$ in terms of the square ratio of the Hubble radius  $H_0^{-1}$ to the physical
length scale of the configuration,
\be
\ep \equiv \left(\frac{H_0^{-1}}{S(t)r_\mathrm{i}}\right)^2
=(\dot{S}r_\mathrm{i})^{-2}
=\frac{t_\mathrm{i}^{2\al}t^{\beta}}{\al^2 r_\mathrm{i}^2},\quad \beta\equiv 2(1-\al).\label{defofep}
\ee
When we set the initial conditions for PBH
formation, the size of the perturbed 
region is much larger than the
Hubble horizon. 
This means $\ep\ll 1$  
at the beginning, so it can serve as an
expansion parameter to construct an analytic solution of 
the system of Einstein equations to describe the spatial dependence of
 all the above variables 
at the initial moment when we set the initial conditions. 
In this paper, the second order solution, obtained in \cite{1475-7516-2012-09-027} (hearafter PNY), is basically used to provide the initial conditions 
for the numerical computations. 


We define the initial curvature profile as
\be
K(0,r)\equiv K_\mathrm{i}(r),
\ee
where $K_\mathrm{i}(r)$ is an arbitrary function of $r$ 
which vanishes outside the perturbed region.
%
%
We normalize radial Lagrangian coordinate $r$ in such a way 
that $K_\mathrm{i}(0)=1$.

In order to represent the comoving length scale of the 
perturbed region,
we use the co-moving radius,
$r_\mathrm{i}$, of the overdense region.
We can calculate $r_{\mathrm i}$ by 
approximately solving (see PNY) the following equation for the energy density perturbation defined by (\ref{defofdelta}):
\be
\delta(t,r_{\mathrm i})=0.\label{defri}
\ee

\section{Two master parameters cruicial for PBHs formation }
\begin{figure}[h]
\begin{center}
\includegraphics[width=16cm,keepaspectratio,clip]{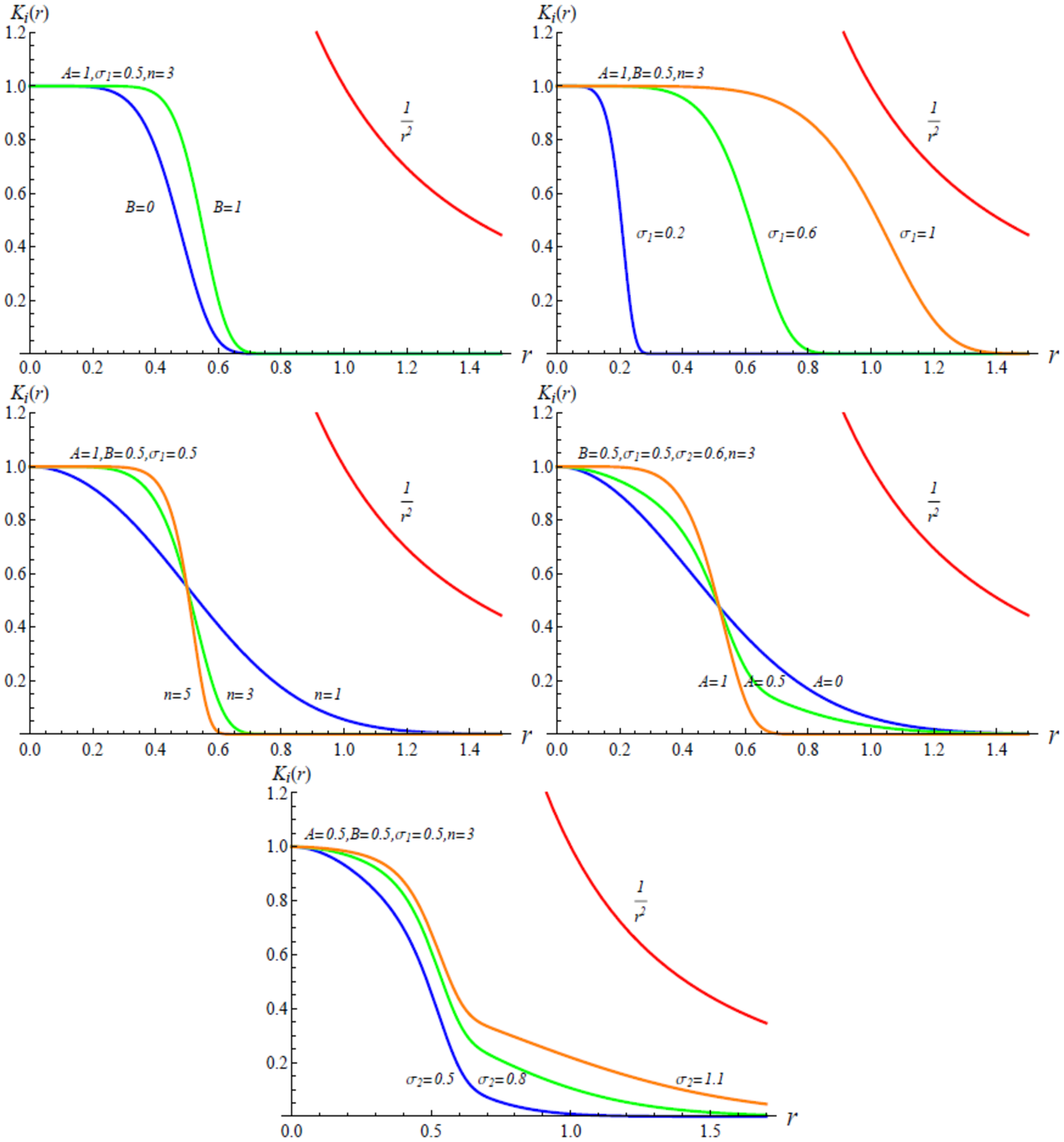}
\end{center}
\caption{
Dependence of the shapes of profiles represented by (\ref{newfunction}) on parameters. 
In each panel, the dependence of the shape on one of the parameters is shown with the rest of the parameters fixed. 
}
\label{shapes608}
\end{figure}

We now proceed to our full analysis (for more details, see \cite{1475-7516-2014-01-037}) 
introducing the following function
\be
K_\mathrm{i}(r)=A\left[1+B\left(\frac{r}{\sigma_1}\right)^{2n}\right]
\exp\left[-\left(\fr{r}{\sigma_1}\right)^{2n}\right]
+(1-A)\exp\left[-\left(\fr{r}{\sigma_2}\right)^{2}\right],\label{newfunction}
\ee
which can represent various shapes of profiles using the five parameters
as is shown in Figure \ref{shapes608}. 
This function not only includes those investigated in previous work but also 
enables us to investigate new shapes of profiles. 

It turned out that a relatively clear separation between configurations which collapse to PBHs and
those
which do not 
is obtained by the following combination:
\be
\Delta\equiv r_{1/6}-r_{5/6}
\ee
and 
\be
I\equiv \int^{r_{3/5}}_0r^2K_\mathrm{i}(r)dr.
\ee

\begin{figure}[h]
\begin{center}
\includegraphics[width=16cm,keepaspectratio,clip]{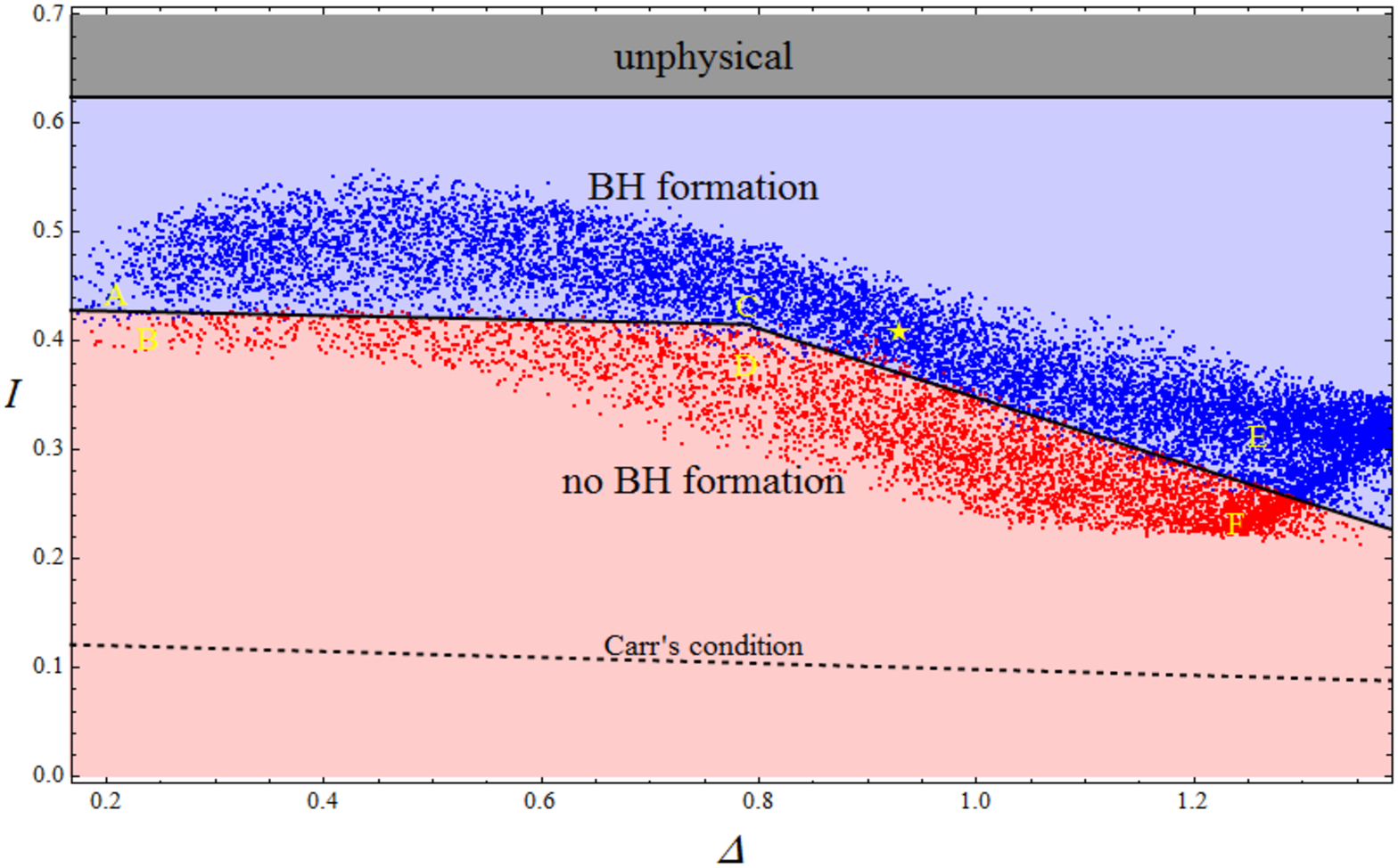}
\end{center}
\caption{
The PBH formation condition for the profiles represented by (\ref{newfunction}). 
The blue and red points correspond respectively to the profiles which lead to the black hole formation and 
those which do not. The shaded region labelled "unphysical" corresponds to the profiles 
which do not satisfy $K_\m{i}(r)<1/r^2$ and therefore are unphysical. 
The profile used as an example of the PBH-forming cases in this paper corresponds to the yellow star in this figure.
The dashed line corresponds to the Carr's condition.
}
\label{conditionink}
\end{figure}

Figure \ref{conditionink} shows the results of numerical calculation
with various initial conditions of the five-parameter family
(\ref{newfunction}). Specifically we have chosen the values of model
parameters in (\ref{newfunction}) in the range $0 \leq A \leq 1$, 
$0 \leq B \leq 1$, $\sigma_\m{min} \leq \sigma_1 \leq \sigma_2 \leq \min\{2\sigma_1,1.14/\sqrt{1-A}\}$,
where $\sigma_\m{min}$ is chosen to search only the profiles relevant to revealing the PBH formation condition 
and
$n=1,2,3,4,5$. 
As is seen there the
condition for PBH formation can be quite well described by the following fitting formula:
\be
(S_1(\Delta-\Delta_{\mathrm{b}})+I_{\mathrm{b}})\Theta(-(\Delta-\Delta_{\mathrm{b}}))
+(S_2(\Delta-\Delta_{\mathrm{b}})+I_{\mathrm{b}})\Theta(\Delta-\Delta_{\mathrm{b}})<I
,\label{fitting}
\ee
where $\Theta$ denotes the unit step function and 
$(S_1,S_2,\Delta_{\rm{b}},I_{\rm{b}})=(-0.021,-0.32,0.79,0.41)$, which represent the slopes of 
the two lines and the position of the break. 
This formula corresponds to the lower solid line in Figure \ref{conditionink}.

Note that for the larger values of $\Delta$, the threshold value for PBH formation $I$ is smaller.
This is  because when $\Delta$ is larger, the pressure gradients are smaller and in addition
gravity is relatively stronger even away from the centre, in which case gravity near the centre, 
measured by $I$, needs not be so large compared to cases with a smaller $\Delta$. 
Put differently, for $I \lesssim 0.43 \equiv I_\m{cr}$, profiles with a smaller
$\Delta$ do not result in PBH formation because the pressure gradient is
so large that the gravitational collapse is hindered.
The dashed line in Figure \ref{conditionink} corresponds to the Carr's condition eq.(\ref{classical}). 

\section{Conclusion}
In this paper we have presented the results of 
numerical computations of the time evolution of a perturbed 
region after the horizon 
re-entry.  The initial conditions for these numerical computations 
were given using an analytical asymptotic expansion technique developed in our previous paper. 
By calculating the time evolution of various initial 
perturbations, the condition for PBH formation has been investigated. 
We have extended preceding analyses by performing many more numerical
computations of PBH formation based on the initial curvature profiles
characterized by five parameters which not only reproduce the variety of 
profiles near the centre but also incorporate the possible extended features
in the tail region (see eq.(\ref{newfunction})). 

We have shown that the criterion of PBHs formation can still 
be expressed in terms of two crucial (master) parameters 
which correspond to the averaged amplitude of over density in the central region and the width of transition region at 
outer boundary. As is shown in Figure \ref{conditionink}, this is
the case even though our profiles are characterized by as many as five parameters.
We have also provided a reliable physical interpretation of the two-parametric 
criterion.


\Acknowledgements
This work was partially
supported by JSPS Grant-in-Aid for Scientific Research 23340058 (J.Y.), 
Grant-in-Aid for Scientific Research on Innovative Areas No. 21111006 (J.Y.),
Grant-in-Aid for Exploratory Research No. 23654082(T.H.),
and Grant-in-Aid for JSPS Fellow No. 25.8199 (T.N.).
TN thanks School of Physics and Astronomy, Queen Mary College, University of London
for hospitality received during this work.
We thank B. J. Carr for useful communications.
TN acknowledges H. Kodama, K. Kohri, K. Ioka and H. Takami for helpful comments.

\bibliography{bib205}
 
\end{document}

%% file: econfmacros.tex
\def\beq{\begin{equation}}
\def\eeq#1{\label{#1}\end{equation}}
\def\eeqn{\end{equation}}

\def\beqa{\begin{eqnarray}}
\def\eeqa#1{\label{#1}\end{eqnarray}}
\def\eeqan{\end{eqnarray}}

\let\bar=\overbar

